\documentclass[journal=jacsat,manuscript=article]{achemso}

\usepackage[T1]{fontenc} % Use modern font encodings
\usepackage{xcolor}% Includes text colors
\usepackage{graphicx}% Include figure files
\usepackage{dcolumn}% Align table columns on decimal point
\usepackage{bm}% bold math
\usepackage{hyperref}% add hypertext capabilities
\setcitestyle{super}
\usepackage{amsmath}
\usepackage{amssymb}
\usepackage[utf8]{inputenc} %for accents

\author{Rajveer Fandan}
\email{rajveer.fandan@upm.es}
\affiliation{Instituto de Sistemas Optoelectrónicos y Microtecnología, Universidad Politécnica de Madrid, Av. Complutense 30, Madrid 28040, Spain}
\alsoaffiliation{Departamento de Ingeniería Electrónica, E.T.S.I de Telecomunicación, Universidad Politécnica de Madrid, Av. Complutense 30, Madrid 28040, Spain}

\author{Jorge Pedrós}
\email{j.pedros@upm.es}
\affiliation{Instituto de Sistemas Optoelectrónicos y Microtecnología, Universidad Politécnica de Madrid, Av. Complutense 30, Madrid 28040, Spain}
\alsoaffiliation{Departamento de Ingeniería Electrónica, E.T.S.I de Telecomunicación, Universidad Politécnica de Madrid, Av. Complutense 30, Madrid 28040, Spain}

\author{Fernando Calle}
\affiliation{Instituto de Sistemas Optoelectrónicos y Microtecnología, Universidad Politécnica de Madrid, Av. Complutense 30, Madrid 28040, Spain}
\alsoaffiliation{Departamento de Ingeniería Electrónica, E.T.S.I de Telecomunicación, Universidad Politécnica de Madrid, Av. Complutense 30, Madrid 28040, Spain}

\title{Exciton-Plasmon Coupling in 2D Semiconductors by Surface Acoustic Waves}% Force line breaks with \\

\begin{document}
\begin{abstract}
We theoretically demonstrate the coupling between excitons in 2D semiconductors and surface plasmons in a thin metal film by means of a surface acoustic wave (SAW), proving that the generated exciton-plasmon polaritons (or plexcitons) are in the strong coupling regime. The strain field of the SAW creates a dynamic diffraction grating providing the momentum match for the surface plasmons, whereas the piezoelectric field, that could dissociate the excitons, is cancelled out by the metal. This is exemplified for monolayer $\mathrm{MoS_{2}}$ and mono- and few-layer black phosphorus on top of a thin silver layer on a $\mathrm{LiNbO_{3}}$ piezoelectric substrate, providing Rabi splittings of 100-150 meV. Thus, we demonstrate that SAWs are powerful tools to modulate the optical properties of supported 2D semiconductors by means of the high-frequency localized deformations tailored by the acoustic transducers, that can serve as electrically switchable launchers of propagating plexcitons suitable for active high-speed nanophotonic applications. 
\end{abstract}

\textbf{Keywords:} Exciton-plasmon polariton, plexciton, strong coupling, surface acoustic wave, $\mathrm{MoS_{2}}$, black phosphorus

\textbf{ORCID iDs of the authors:}

Rajveer Fandan: http://orcid.org/0000-0002-4885-8853

Jorge Pedrós: https://orcid.org/0000-0002-3154-0187

Fernando Calle: https://orcid.org/0000-0001-7869-6704

\section*{Introduction}

2D semiconductors, such as transition metal dichalcogenides (TMDCs) and phosphorene/ black phosphorus (BP), have attracted a lot of interest in photonics due to their strong excitonic effects and valley-dependent properties. \cite{mak2016photonics,ling2015renaissance}. Because of their reduced dimensionality, 2D semiconductors present stronger confinement and reduced screening as compared to 3D semiconductors, leading to enhanced Coulomb interactions. This makes electron-hole quasiparticles, such as excitons, charged excitons (trions), and biexcitons, observable even at room temperature due to their high binding energies. Their reduced dimensionality also leads to more efficient light absorption and to stronger light-matter interactions, making them very sensitive to the environment. Thus, excitons in 2D semiconductors, covering the visible and near infrared ranges, can couple very efficiently to surface plasmons in metals, leading to exciton-plasmon polaritons (or plexcitons) in the strong coupling regime, where the rate of energy transfer between the exciton and the plasmon exceeds the total damping rate of the system. Moreover, excitons in 2D semiconductors present the unique feature of the valley degree of freedom, i.e. optical transitions in distinct valleys in momentum space. This valley polarization of excitons has been conﬁrmed to be retained by the polaritonic excitations under strong-coupling conditions \cite{Lundt_2017, Dufferwiel2017}, paving the way for future valleytronic architectures \cite{Schneider2018}. 

For enabling the exciton-plasmon strong coupling, individual metallic nanostructures can simply be placed on top of the 2D semiconductor \cite{Wen2017,Cuadra2018} to host localized surface plasmon resonances (LSPRs). Higher quality factors can be achieved, though, when metal is nanostructured into periodic arrays, where LSPRs couple coherently via the Rayleigh’s anomaly, providing narrower resonances. The 2D semiconductors are then transferred onto this plasmonic lattice \cite{liu2016strong}. However, continuous metal architectures hosting propagating plasmons are desired for implementing devices where the exciton-plasmon polaritons can serve as ultra-fast and low-power signal carriers. Metal nanowires \cite{Goodfellow:14,Lee2016,Kim2019} and films with patterned gratings \cite{Klein2019} have been used for launching plasmons that propagate along the metal and into the 2D semiconductors, coupling to their excitons, demonstrating various nanophotonic device and circuit functionalities.

Gratings offer a simple on-chip means to overcome the wave-vector mismatch and to convert far-field light into propagating plasmons. However, they are intrinsically static, thus lacking the tunability required in modern high-speed nanophotonics. Surface acoustic waves (SAWs) have been shown to produce dynamic diffraction gratings capable of launching propagating plasmons in metal films \cite{sun1991interactions,ruppert2010surface} and graphene \cite{PhysRevLett.111.237405}, that can be switched electrically.

In this paper, we demonstrate that the exciton-plasmon strong coupling might also be facilitated in a 2D semiconductor onto an unpatterned thin metal layer by using the dynamic strain field of SAW, leading to propagating plexcitons. In the proposed approach, a transducer on a piezoelectric substrate hosting a metal/2D semiconductor system is used to launch a SAW. The SAW modulates the surface forming a dynamic diffraction grating that drives the generation of plasmons in the metal film, which can then couple with the optically generated excitons in the 2D semiconductor atop. This mechanism has been studied for various direct band gap 2D semiconductors, where excitons are especially intense, including monolayer TMDCs and mono- and few-layer BP. The dissociation of the excitons, produced by the piezoelectric field accompanying the SAW in the case of piezoelectric 2D semiconductors \cite{Rezk2015, Rezk2016}, which is detrimental for the formation of the polaritons, is cancelled out by the metal film itself, in the same way as the piezoelectric field coming from the piezoelectric substrate. The design parameters for the distinct 2D material systems are discussed in detail.

\section*{Experimental Details}
\begin{figure}
    \includegraphics[width=0.5\textwidth,keepaspectratio]{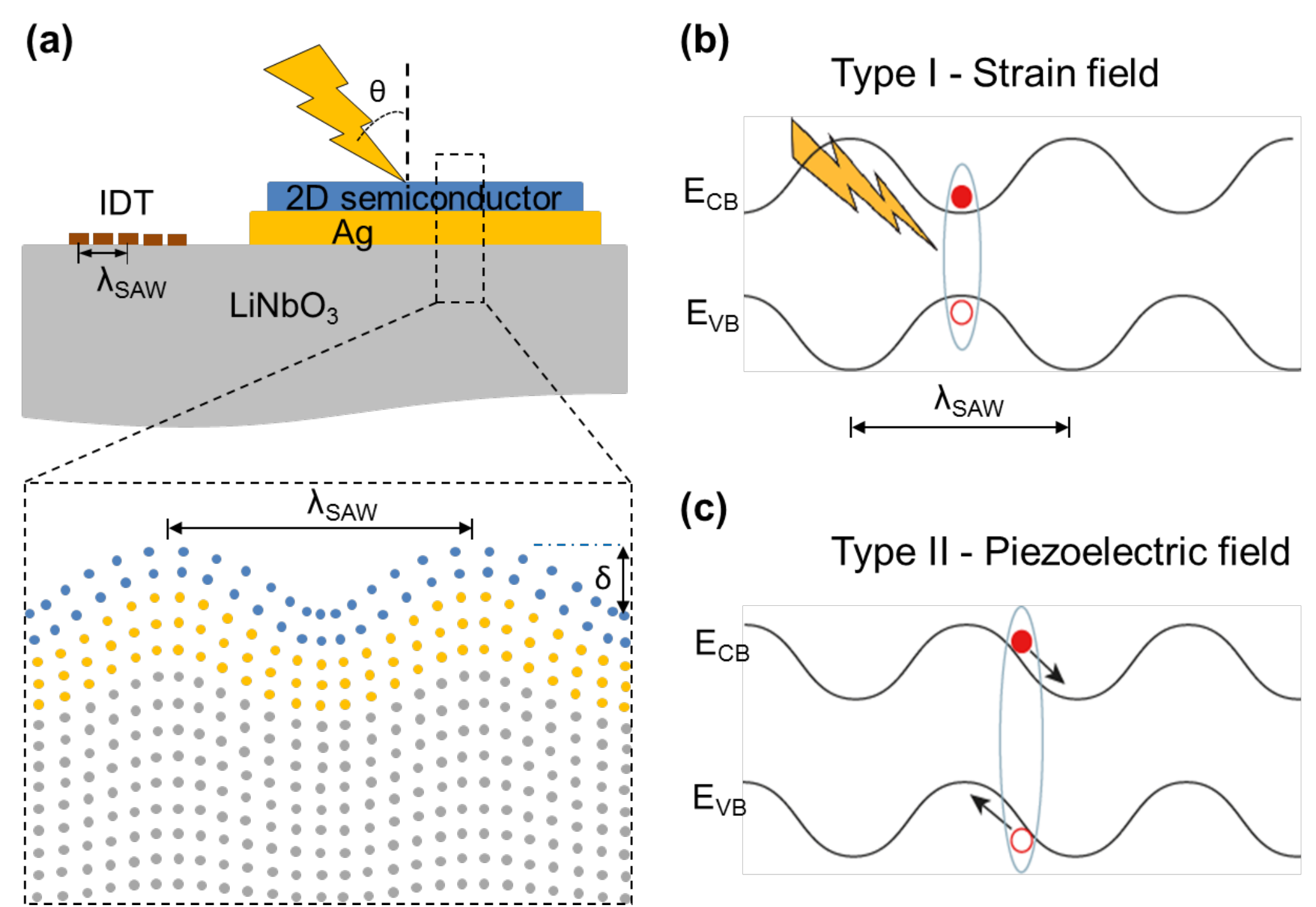}
    \caption{Schematics of the (a) device proposed and the (b) energy band diagram modulation of the 2D semiconductor produced by the strain and piezoelectric fields of the SAW, where the exciton formation is indicated. The inset in (a) shows the profile of the SAW propagating across the 2D semiconductor/Ag/$\mathrm{LiNbO_{3}}$ stack in the sagittal plane. $\lambda_{SAW}$ and $\delta$ denote the wavelength and the amplitude of the SAW.}
    \label{fig:5.1}
\end{figure}

A stack geometry of 2D crystal ($\mathrm{MoS_{2}}$, BP)/Ag/$\mathrm{LiNbO_{3}}$ has been considered for the proposed device, as shown schematically in Figure \ref{fig:5.1}(a). Silver (Ag) has been chosen as the plasmonic metal because its mass density is smaller than that of gold, the other typically employed metal for plasmonics, since mass loading damps the SAW amplitude. And $\mathrm{LiNbO_{3}}$ has been selected as the piezoelectric substrate due to its large electromechanical coupling coefficient, that ensures a strong SAW that modulates efficiently the 2D material on top \cite{Fandan2020}. Thus, an interdigital transducer (IDT) fabricated on the $\mathrm{LiNbO_{3}}$ substrate and fed with a high-frequency RF signal launches a SAW that modulates periodically the surface of the substrate and the whole 2D crystal/Ag stack on top of it. 

The sinusoidal deformation of the surface produced by the SAW creates a virtual dynamic diffraction grating that provides the extra wave vector required for the light to couple into plasmons in the Ag layer, that hybridize with the excitons in the 2D semiconductor. Thus, in the presence of a SAW of wavelength $\lambda_{SAW}$ and amplitude $\delta$, the propagation of TM-polarized light is reduced as the SAW-induced diffraction grating scatters the incident light with wave vector ($q_{\vert\vert,0}$, $q_{z,0}$) into the various diffraction orders ($q_{\vert\vert,m}$, $q_{z,m}$) with 
\begin{equation}
      q_{\vert\vert,m}=\dfrac{\omega}{c}sin\theta+m\dfrac{2\pi}{\lambda_{SAW}},
\end{equation}
where  $\theta$  is  the  angle  of  off-normal  incidence  and  $m$ is  an integer. The wavelength of the plasmon $\lambda$ is hence mainly governed by $\lambda_{SAW}$, as $(\omega/c)sin\theta << 2\pi/\lambda_{SAW}$.

The deformation of the lattice produced by the SAW also leads to a modulation of the band diagram \cite{Lima_2005} of the 2D semiconductor, affecting its excitonic behaviour. Both, the strain and the piezoelectric fields of the SAW intervene in this modulation, as depicted in Figure \ref{fig:5.1}(b). The strain field induces an anti-symmetric (type-I) band diagram modulation, with the minimum and maximum band gap occurring at the regions of maximum tension and compression, respectively. The piezoelectric field, on the contrary, induces a symmetric (type-II) band diagram modulation with constant band gap, that ionizes the excitons and spatially separates the free electrons and holes. Therefore, the thickness of the metal underneath the 2D semiconductor should be chosen in such a way that it completely screens both the piezoelectric field accompanying the SAW generated in the piezoelectric substrate and that of the 2D crystal itself, if it is also piezoelectric. In particular, $\mathrm{MoS_{2}}$ and other TMDCs, with an odd number of layers, and BP, with both odd and even number of layers, present in-plane piezoelectricity due to their non-centrosymmetric structure \cite{Cui2018,Hinchet2018}. For Ag, the minimum thickness required to completely screen an electric field has been reported to be 5 nm \cite{fahy1988electromagnetic,schad1992metallic}. 

A transfer matrix method (TMM) \cite{Fandan_2018} has been used to calculate the exciton-plasmon polariton dispersion in the layered system. The frequency-dependent optical response of Ag is given by its relative dielectric function, which can be expressed in Drude model form
\cite{yang2015optical,abd2019plasmonics} as
\begin{equation}
    \epsilon_{Ag}(\omega)=\epsilon_{\infty}-\dfrac{\omega_{p}^2}{\omega(\omega+i\gamma_{Ag})},
\end{equation}
where $\epsilon_{\infty}=3$ is the high frequency dielectric constant, $\omega_{p}=9.1$ eV is the plasma frequency and $\gamma_{Ag}=20$ meV is the damping rate. In the optical range of interest, the effect of the optical phonons of $\mathrm{LiNbO_{3}}$ can be neglected and the relative dielectric function of the material can be reduced to just its dielectric constant $\epsilon_{LiNbO_{3}}=5$ \cite{sun1991interactions}.

The frequency-dependent optical response of the 2D materials can be described by their 2D conductivity as \cite{ciuti1998role,Yu2010,Soh_2017}
\begin{equation}
    \sigma(\omega)=\dfrac{-2i\omega \left|\Psi\right|^2 \langle\left|e.d\right|^2\rangle}{E_{ex}-\omega-i\gamma_{ex}},
\end{equation}
where $\Psi$ is the exciton wave function, $e$ is the photon polarization, $d$ is the transition dipole moment matrix element, $E_{ex}$ is the exciton energy and $\gamma_{ex}$ is the exciton linewidth taken as $\mathrm{1 \hspace{1mm}ps^{-1}}$ (=  0.66 meV). For reaching the strong exciton-plasmon coupling regime, the Rabi splitting ($\hbar\omega_{R}$) must exceed the total damping rate, i.e. $\hbar\omega_{R}>\gamma_{ex}+\gamma_{Ag}$. 

The results of the TMM computation have been compared to those obtained using a simple coupled oscillator model (COM), which allows us to gain extra insight into the physics of the coupling. The exciton-plasmon polariton dispersion in the COM is described by the following system of equations:
\begin{equation}
     \begin{bmatrix}
       E_{ex}+i\gamma_{ex} & g           \\[0.3em]
       g & E_{pl}(q)+i\gamma_{Ag} \\[0.3em]
     \end{bmatrix}
     \begin{bmatrix}
       X           \\[0.3em]
       C \\[0.3em]
     \end{bmatrix}=E\begin{bmatrix}
       X           \\[0.3em]
       C \\[0.3em]
     \end{bmatrix},
\end{equation}
where $E_{ex}$ is the exciton energy, $E_{pl}(q)$ is the $q$-dependent plasmon resonance energy (calculated in the Ag/$\mathrm{LiNbO_{3}}$ system by TMM), $g$ is the exciton-plasmon coupling strength, $E$ is the eigenvalue corresponding to the energies of the polariton modes, and $X$ and $C$ are the components of the eigenvectors. The latter fulfill the condition $\vert X \vert^{2} + \vert C \vert^{2} =1$, where $\vert X \vert^{2} $ and $ \vert C \vert^{2}$ are referred to as the Hopfield coefficients, that represent weighing fractions of the exciton and plasmon contribution to the polaritons, respectively. Note that this condition is fulfilled at time zero and that the system will evolve with time as the polariton gets damped. The analytical solutions of the system of equations are the two polariton branches given by the expression
\begin{equation}
    E_{\pm} =\dfrac{E_{ex}+E_{pl}(q)}{2} +\dfrac{i(\gamma_{ex}+\gamma_{Ag})}{2}\pm \dfrac{\sqrt{4g^2+[E_{ex}-E_{pl}(q)+i(\gamma_{ex}-\gamma_{Ag})]^2}}{2},
\end{equation}
where the $\pm$ sign in the eigenvalue E corresponds to the upper polariton (UP) and lower polariton (LP), respectively, $g=\sqrt{(E_{ex}-E_{-})(E_{+}-E_{ex})}$ is the coupling strength and $E_{pl}(q)-E_{ex}$ is the detuning between the plasmon resonance and the exciton transition energy.
The lifetime of the polaritons is directly determined by the Hopfield coefficients and $\gamma_{ex}$, $\gamma_{Ag}$ as:
\begin{equation}
    \gamma_{LP} =\vert X \vert^{2}\gamma_{ex}+\vert C \vert^{2}\gamma_{Ag}
\end{equation}    
\begin{equation}
    \gamma_{UP} =\vert C \vert^{2}\gamma_{ex}+\vert X \vert^{2}\gamma_{Ag}.
\end{equation}

\section*{Results and Discussion}

\begin{figure}
\includegraphics[width=0.5\textwidth,keepaspectratio]{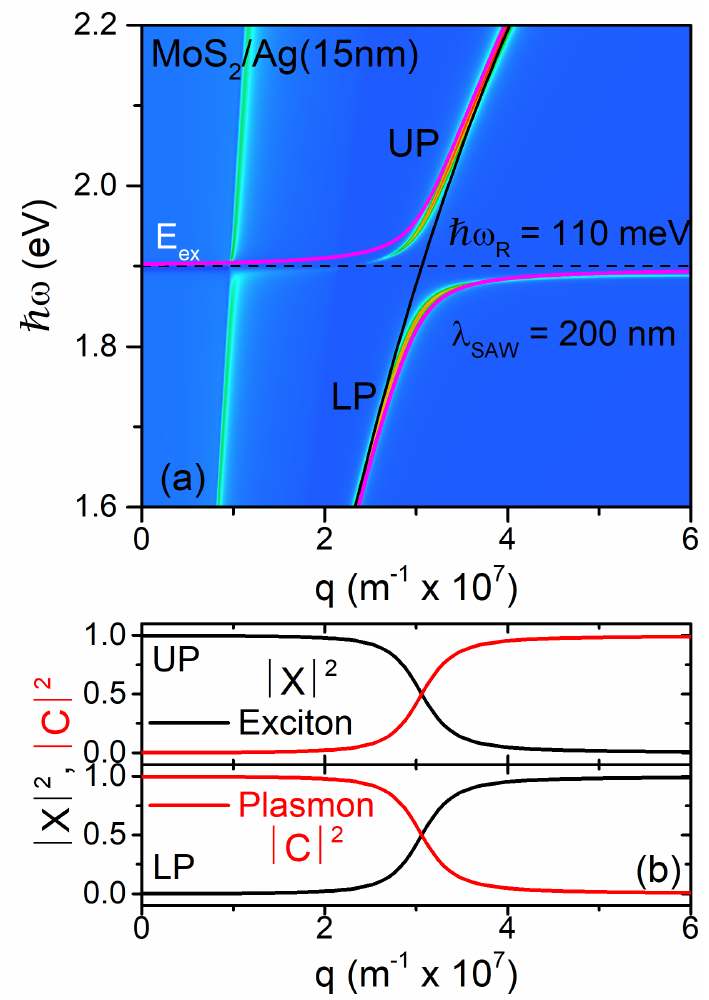}
\caption{(a) Exciton-plasmon polariton dispersion for a $\mathrm{MoS_{2}}$/Ag/$\mathrm{LiNbO_{3}}$ system with a 15 nm-thick Ag layer. UP and LP denote the upper polariton (UP) and lower polariton (LP) branches, respectively. The contour plot corresponds to the TMM computation, whereas the pink curves are the COM eigenvalues. The solid black line is the energy dispersion of the Ag plasmon, $E_{pl}(q)$, whereas the dashed black line indicates the exciton energy, $E_{ex}$. The values of the Rabi splitting, $\hbar\omega_{R}$, and the SAW wavelength, $\lambda_{SAW}$, are also indicated.
(b) Hopfield coefficients, $\vert X \vert^{2} $ and $ \vert C \vert^{2}$, of the UP and LP branches in (a) obtained in the COM description, indicating the relative exciton and plasmon contribution to the polaritons.}
\label{fig:5.2}
\end{figure}

We consider first a $\mathrm{MoS_{2}}$/Ag/$\mathrm{LiNbO_{3}}$ system. Figure \ref{fig:5.2}(a) shows the calculated exciton-plasmon polariton dispersion for this system with a 15 nm-thick Ag layer. The exciton energy of $\mathrm{MoS_{2}}$ is $E_{ex}$=1.9 eV \cite{li2014measurement} (dashed line), whereas the plasmon energy $E_{pl}(q)$ (black line)
is already dispersive. The exciton-plasmon coupling leads to two polaritonic branches, denoted as UP and LP, that result from the anti-crossing of the exciton and plasmon curves. A Rabi splitting of 110 meV is produced at a $q$ value corresponding to a SAW wavelength of 200 nm ($q=2\pi/\lambda_{SAW}$). The contour plot corresponds to the TMM computation, that matches consistently the COM eigenvalues represented by the pink curves. Figure \ref{fig:5.2}(b) presents the Hopfield coefficients of the UP and LP branches calculated in the COM description. For $q$ = 0, the UP is purely excitonic in character ($\vert X \vert^{2} $ = 1 and $ \vert C \vert^{2}$ = 0) and the LP is purely plasmonic ($\vert X \vert^{2} $ = 0 and $ \vert C \vert^{2}$ = 1), whereas the opposite situation occurs for high $q$ values. At the anticrossing point (for $q=2\pi/\lambda_{SAW}$), where the Rabi splitting is defined, the two quasiparticles are hybridized, being half exciton-half plasmon ($\vert X \vert^{2} $ = $ \vert C \vert^{2}$ = 0.5).

Thinner metal layers lead to larger values of the Rabi splitting but at higher $q$ values. This is due to the fact that decreasing the metal thickness moves the plasmon dispersion, $E_{pl}(q)$, further away from the light line (light cone) towards a larger value of $q$, thus a smaller $\lambda$. The latter implies a larger electric field density and a higher proportion of this electric field distribution lying in the vicinity of the 2D excitonic medium (i.e., the spatial overlap between the fields of the plasmon and the exciton increases). The Rabi splitting also depends on the coupling strength, $g$, which can be enhanced by using emitters with a large dipole moment (i.e., a large oscillator strength) and/or by enhancing the electric field of the plasmonic cavity mode (i.e. a plasmon with smaller $\lambda$). The latter is proportional to $1/\sqrt{V}$, where $V$ is the mode volume, which for plasmonic nanocavities goes as $\sim \lambda^3$ \cite{marquier2017revisiting,akselrod2015leveraging}. Therefore, confining the electromagnetic energy in a small $V$ enhances the interaction and thus leads to a large Rabi splitting. This poses a major challenge for the current SAW technology in achieving very small $\lambda_{SAW}$ as, nowadays, the state-of-the-art limit with e-beam and nano-imprint lithographies is of the order of $\lambda_{SAW}$ = 160 nm \cite{kukushkin2004ultrahigh,buyukkose2012ultrahigh,buyukkose2013ultrahigh,zheng202030}. One way to overcome this limitation is to increase slightly the metal thickness so that the plasmon dispersion shifts towards the light line and hence the maxima of the exciton-plasmon splitting occur at smaller $q$ values. Although this is at the expense of a reduction in the Rabi splitting, this is still large enough to be in the strong coupling regime.

\begin{figure}
\includegraphics[width=0.5\textwidth,keepaspectratio]{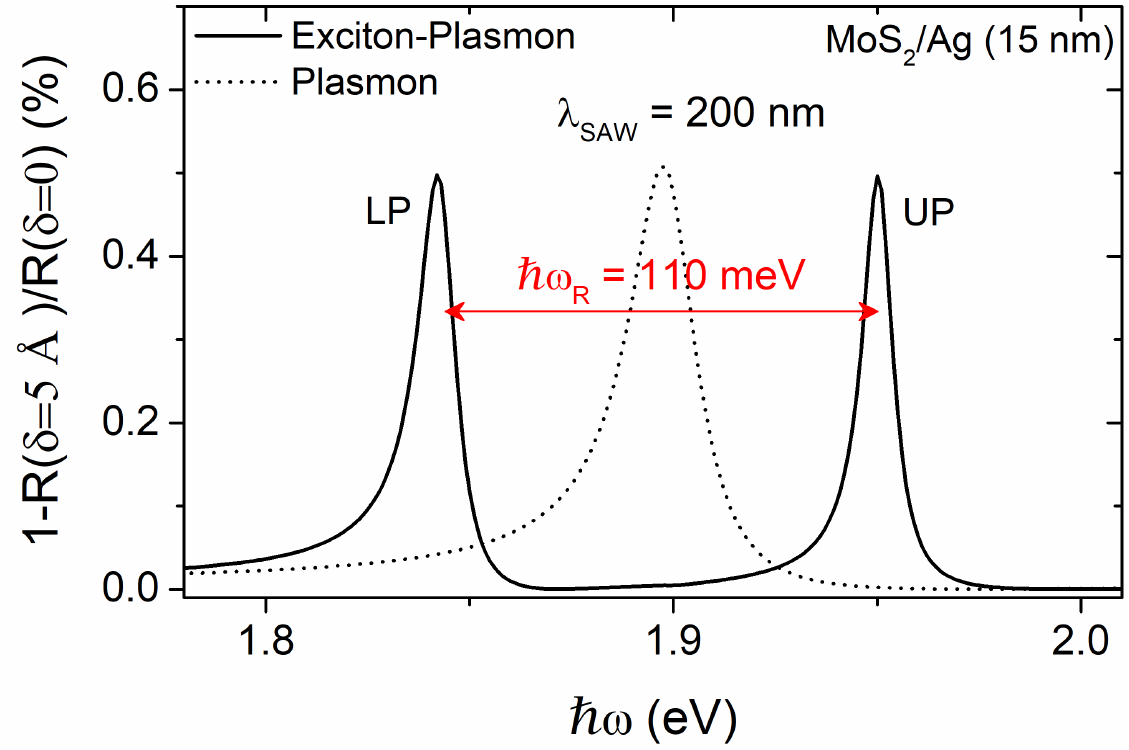}
\caption{Change in the reflectance spectra induced by a SAW of amplitude $\delta=$ 5 {\AA} and wavelegth $\lambda_{SAW}$ = 200 nm in a $\mathrm{MoS_{2}}$/Ag/$\mathrm{LiNbO_{3}}$ system with a 15 nm-thick Ag layer. The single peak in the dashed curve is the reflectance of the bare Ag plasmon (i.e., without the exciton coupling), whereas the splitting of the reflectance into two peaks in the solid curve corresponds to the upper (UP) and lower (LP) polariton branches produced by the the Ag plasmon coupled to the $\mathrm{MoS_{2}}$ exciton. The value of the Rabi splitting $\hbar\omega_{R}$ is indicated.
}
\label{fig:5.3}
\end{figure}

The exciton-plasmon coupling leads to an enhanced absorption of the light, which can be detected easily as a change in the reflectance spectrum of the material system when the SAW is used to generate plasmons in the metal layer that couple to excitons in the 2D semiconductor. Figure \ref{fig:5.3} plots the differential reflectance or extinction spectra of the $\mathrm{MoS_{2}}$/Ag/$\mathrm{LiNbO_{3}}$ system with a 15 nm-thick Ag layer modulated by a SAW with $\delta$ = 5 {\AA} and $\lambda_{SAW}$ = 200 nm, corresponding to the $q$ value at which the Rabi splitting ($\hbar\omega_{R}$) of the system occurs as shown in Figure \ref{fig:5.2}. The solid curve in Figure \ref{fig:5.3} corresponds to the coupling of plasmons and excitons and the separation between the two peaks is a direct measurement of the Rabi splitting, whereas the dashed curve, with a single peak centered at $E_{ex}$, corresponds to the generation of metal plasmons only. The obtained values of $\hbar\omega_{R}$ are comparable with the reported values in the literature \cite{gonccalves2018plasmon}.

Here, we have restricted the discussion to the main exciton of $\mathrm{MoS_{2}}$, exciton A at $E_{ex_A}$ = 1.9 eV, as considering also exciton B at at $E_{ex_B}$ = 2.1 eV \cite{li2014measurement} would only produce an extra splitting, leading to three polaritonic branches, without altering the physics described. 

\begin{figure}
\includegraphics[width=0.5\textwidth,keepaspectratio]{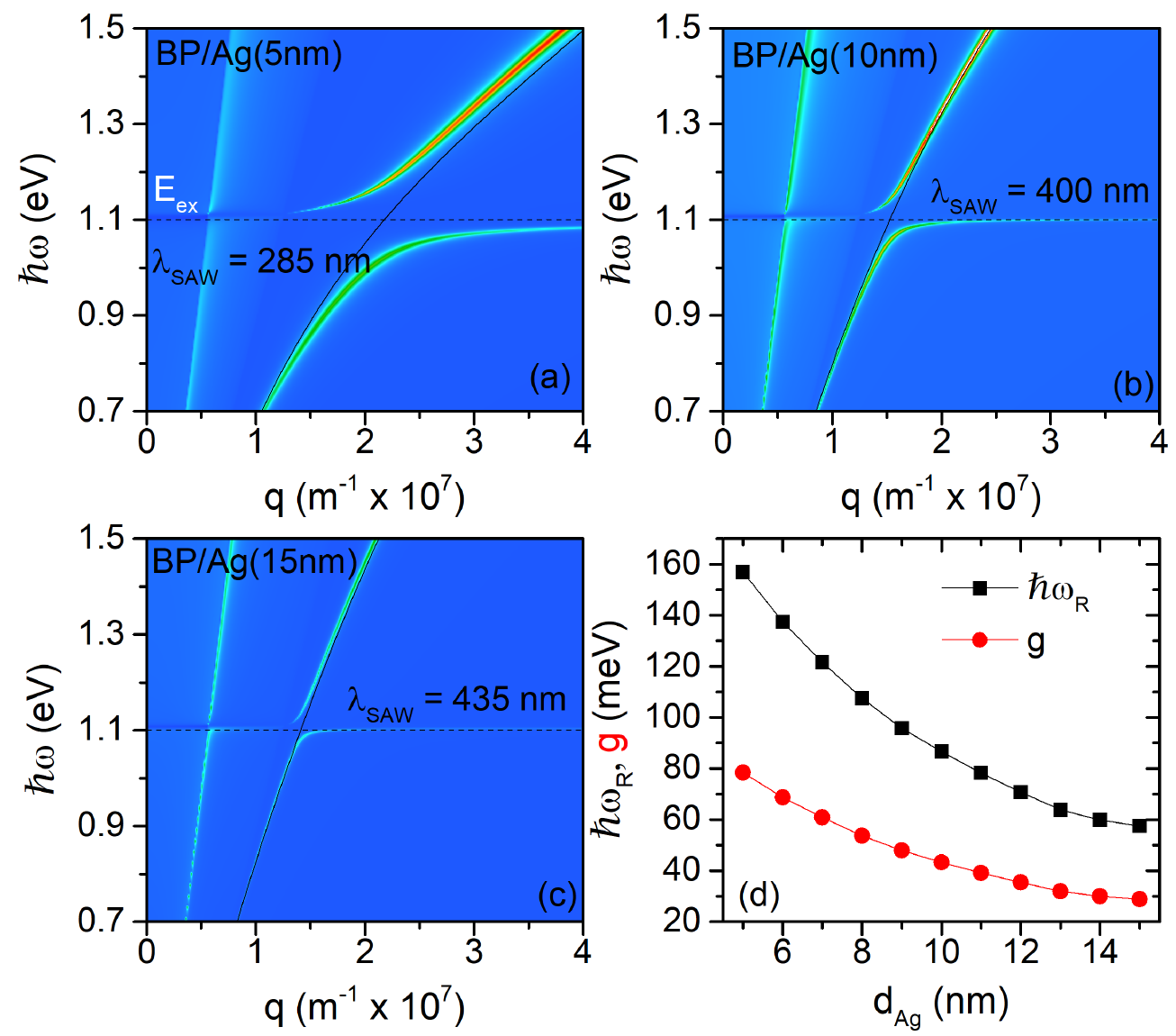}
\caption{Exciton-plasmon dispersion for the  ML BP/Ag/$\mathrm{LiNbO_{3}}$ system with a Ag layer of a thickness of (a) 5 nm, (b) 10 nm and (c) 15 nm. The SAW wavelength, $\lambda_{SAW}$, at which the Rabi splitting occurs is indicated. (d) Variation in the Rabi splitting, $\hbar\omega_{R}$, and the coupling strength, $g$, as a function of the Ag layer thickness.}
\label{fig:5.4}
\end{figure}

$\mathrm{MoS_{2}}$ is a direct band gap semiconductor only in the monolayer (ML) case, otherwise becoming an indirect band gap material. This direct-to-indirect band gap transition is a common feature of most TMDCs. However, phosphorene or black phosphorus is a 2D semiconductor that has a direct band gap for both the ML and few-layer cases, although the value of the band gap energy is layer dependent, decreasing as the number of layers increases. Thus, we consider also here the BP/Ag/$\mathrm{LiNbO_{3}}$ system with a varying number of BP MLs. The exciton energy for ML BP is $E_{ex}$ = 1.1 eV \cite{zhang2018determination}, allowing the exciton-plasmon coupling to occur at lower $q$ values (i.e. larger $\lambda_{SAW}$), as compared to the $\mathrm{MoS_{2}}$ case. Hence, the BP/Ag/$\mathrm{LiNbO_{3}}$ system permits to reduce the Ag thickness down to the minimum of 5 nm required for the screening of the piezoelectric fields, without compromising the limit of $\lambda_{SAW}$ imposed by the resolution of the nanolithography of the IDTs. 
The exciton-plasmon dispersion for the ML BP/Ag/$\mathrm{LiNbO_{3}}$ system with Ag layers of different thickness is plotted in Figure \ref{fig:5.4}(a)-(c). The reduction of the metal layer thickness enhances the strong coupling as shown by the increment in the Rabi splitting, as depicted in Figure \ref{fig:5.4}(d), along with the coupling strength, $g$, which is related to $\hbar\omega_{R}$ as
\begin{equation}
    \hbar\omega_{R}=\sqrt{4g^2-(\gamma_{ex}-\gamma_{Ag})^2}.
\end{equation}

\begin{figure}
\includegraphics[width=0.5\textwidth,keepaspectratio]{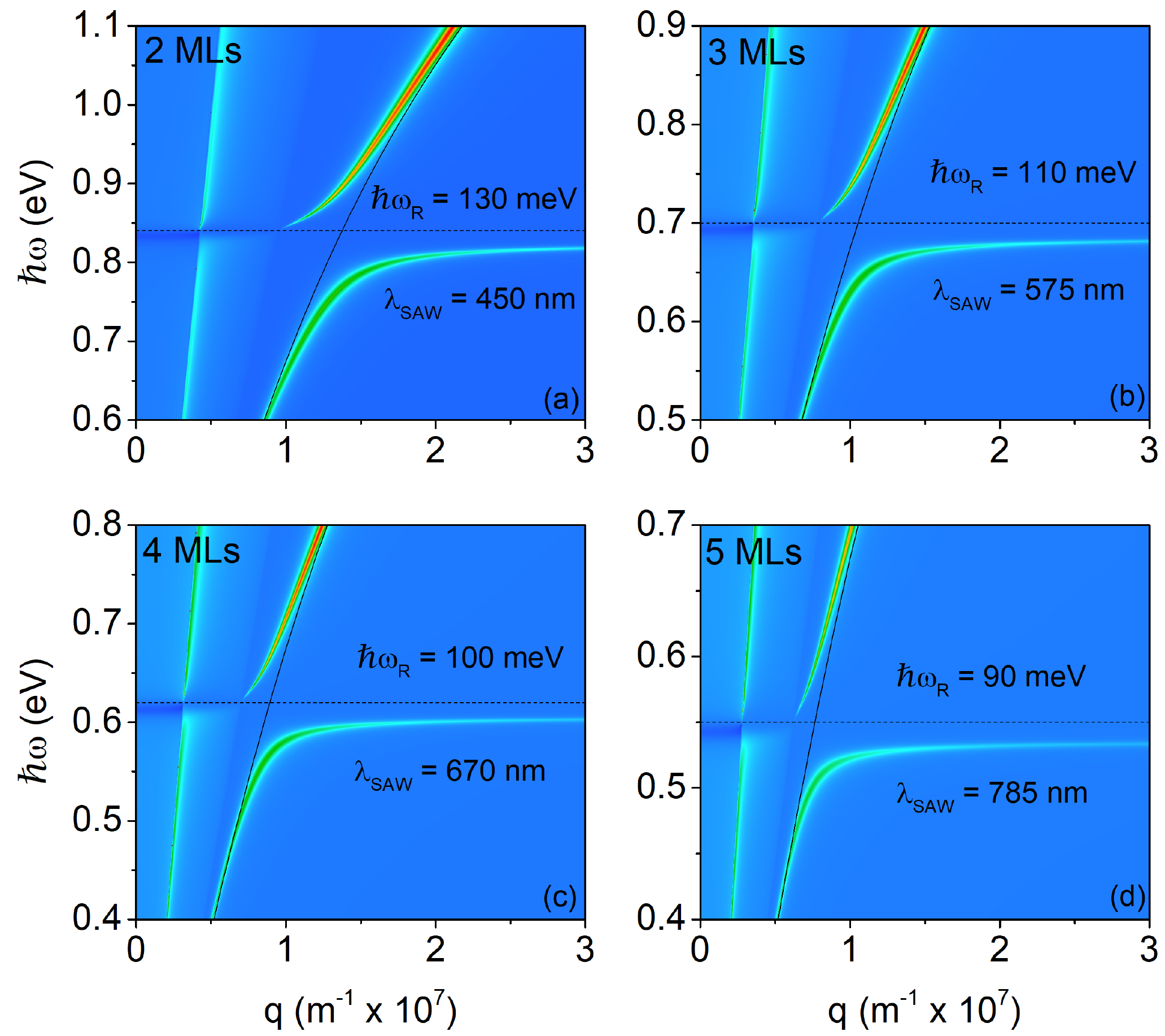}
\caption{Exciton-Plasmon dispersion for the few-layer BP/Ag/$\mathrm{LiNbO_{3}}$ system with a 5 nm-thick Ag layer and 
varying BP monolayer (ML) number, (a)-(d) from 2 to 5 MLs. The SAW wavelength, $\lambda_{SAW}$, at which the Rabi splitting occurs is indicated.
}
\label{fig:5.5}
\end{figure}

Since BP remains direct band gap even with few layers, the few-layer BP/Ag/$\mathrm{LiNbO_{3}}$ system can also host excitons with a layer-dependent exciton energy. Thus, Figure \ref{fig:5.5}(a)-(d) shows the exciton-plasmon dispersion for the BP/Ag/$\mathrm{LiNbO_{3}}$ system, where BP is 2 to 5 ML thick and the Ag layer is 5 nm thick in all cases. $E_{ex}$ decreases from 0.84 to 0.55 eV when moving from 2 to 5 ML-thick BP \cite{zhang2018determination}. The increase in the ML number shifts the Rabi splitting to smaller $q$ values, which can ease the fabrication of the SAW devices while the system remains in the strong coupling regime as $\hbar\omega_{R}$ only reduces slightly (from 130 to 90 meV when moving from 2 to 5 MLs).

\section*{Conclusions}
We have theoretically proven that SAWs can be used to couple optically generated excitons in 2D semiconductors (for example monolayer $\mathrm{MoS_{2}}$ and mono- and few-layer BP) and plasmons in a metal film (Ag) in a ($\mathrm{MoS_{2}}$, BP)/Ag/$\mathrm{LiNbO_{3}}$ stack geometry. The thickness of the Ag layer has been chosen so that it completely screens the piezoelectric field of both the SAW generated in the piezoelectric substrate underneath and the 2D semiconductors above, which are also inherently piezoelectric and can ionize the excitons, leaving behind only the SAW strain field. This coupling leads to two polaritonic branches with large enough Rabi splitting driving these exciton-plasmon systems well into to the strong coupling regime. This Rabi splitting can be increased (decreased) by increasing (decreasing) the thickness of the Ag layer and occurs at large (small) SAW wavevector, which is a design parameter of the SAW device. In the case of BP, which remains a direct band gap semiconductor up to few layers, in contrast with $\mathrm{MoS_{2}}$ which only presents a direct band gap in the monolayer case, the Rabi splitting decreases with increasing the layer number and occurs at larger SAW wavevector. The proposed exciton-plasmon coupling mechanism can be extended to any other direct band gap 2D semiconductor, which can be easily transferred onto a piezoelectric substrate that can generate a SAW. Thus, we have demonstrated that SAWs are powerful tools to mediate strong light-matter interactions in supported 2D semiconductors by means of the high-frequency localized deformations tailored by the acoustic transducers, that can serve as electrically switchable launchers of propagating plexcitons.

Moreover, the obtained results pave the way for more complicated studies like the trapping of polaritons in the moving potential created by the band edge modulation produced by the SAW \cite{Lima06}. The band edge modulation by the SAW piezoelectric and strain fields can induce longitudinal forces on the carriers which are proportional to $dE_{i}/dx$, where $E_{i}$ is the modulated band edge profile (with constant and varying band gap energy in the piezoelectric and strain field cases, respectively), and $x$ is the SAW propagation direction \cite{Sogawa01}. The longitudinal force produced by the SAW strain field for frequencies below 1 GHz ($\lambda_{SAW}$ of several microns) is negligible as compared to that due to the SAW piezoelectric field. However, in our study, since the piezoelectric field is cancelled out and the operating SAW frequencies are of several GHz (sub-micron $\lambda_{SAW}$ regime), this force would be large enough to transport efficiently the exciton-plasmon polaritons over several hundreds of microns with reduced shape distortion. In addition, this approach could also be used to generate SAW-driven polariton condensates, as demonstrated in GaAs-based quantum well (QW) systems within microcavities at cryogenic temperatures \cite{Cerda-Mendez10,PhysRevB.86.100301}, but taking advantage of the fabrication easiness and the tunability of 2D semiconductors. A similar SAW-mediated approach has also been proposed in GaAs-based QW systems with metal stripes on top leading to Tamm plasmon-exciton polaritons \cite{buller2016dynamical}, as a mechanism for signal modulation in polaritonic devices. Here, 2D semiconductors would provide a thermal advantage, as they support Tamm plasmon-exciton polaritons at room temperature \cite{Lundt2016}. On the other hand, the SAW-mediated plasmon-exciton coupling could also be used as a spectroscopy tool to brighten dark excitons, in a similar way as they have been probed using nanoscale trenches etched in a Ag film supporting TMDCs \cite{Zhou2017}.

\begin{acknowledgement}
The authors thank Paulo V. Santos and Francisco Guinea for helpful discussions. This work has received funding from the European Union’s Horizon 2020 Research and Innovation Programme under Marie Skłodowska-Curie Grant Agreement No 642688, from the Spanish Ministry of Science and Innovation (MICINN) through project DIGRAFEN (ENE2017-88065-C2-1-R), from the Comunidad de Madrid through project NMAT2D-CM (P2018/NMT-4511), and from the Universidad Politécnica de Madrid (UPM) through project GRAPOL (VJIDOCUPM18JPA). J.P. acknowledges ﬁnancial support from MICINN (Grant RyC-2015-18968) and UPM (Ayuda para la incorporación y retención de talento doctor, Programa Propio de I+D+i 2019 y 2020).
\end{acknowledgement}

\bibliography{ref}

\clearpage 
\begin{figure}
\includegraphics[width=8.25cm,height=4.45cm]{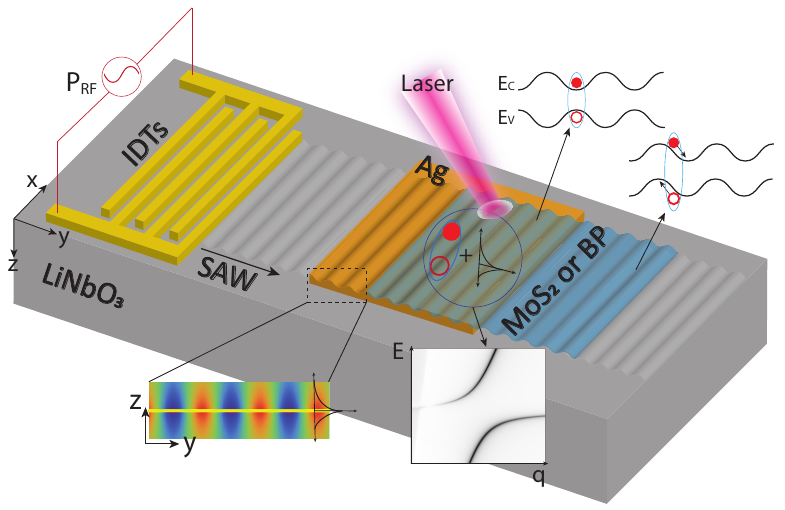}
\caption*{TOC Graphic.}
\end{figure}

\end{document}